# FeFET-based MirrorBit cell for High-density NVM storage

Paritosh Meihar, *Student Member, IEEE*, Rowtu Srinu, Vivek Saraswat, Sandip Lashkare, Halid Mulaosmanovic, *Senior Member, IEEE*, Ajay Kumar Singh, Stefan Dünkel, Sven Beyer and Udayan Ganguly, *Senior Member, IEEE*

*Abstract* — $HfO_2$-based Ferroelectric field-effect transistor (FeFET) has become a center of attraction for non-volatile memory application because of their low power, fast switching speed, high scalability, and CMOS compatibility. In this work, we show an n-channel FeFET-based Multibit memory, termed "MirrorBit", which effectively doubles the chip density via programming the gradient ferroelectric polarizations in the gate, using an appropriate biasing scheme. We have experimentally demonstrated MirrorBit on GlobalFoundries' $HfO_2$-based FeFET devices fabricated at 28 nm bulk HKMG CMOS technology. Retention of MirrorBit states has been shown up to $10^5$ s at different temperatures. Also, the endurance is found to be more than $10^3$ cycles. A TCAD simulation is also presented to explain the origin and working of MirrorBit states, based on the FeFET model calibrated using the GlobalFoundries FeFET device. We have also proposed the array level implementation and sensing methodology of the MirrorBit memory. Thus, we have converted 1-bit FeFET into 2-bit FeFET using programming and reading schemes in existing FeFET, without the need for any special fabrication process alteration, to double the chip density for high-density non-volatile memory storage.

## I. INTRODUCTION

MOORE'S law has been predicting the physical transistor scaling for nearly 6 decades [1]. The requirement for high-density and high-performance memory cells has become essential due to big data, neural network training, IoT, etc [2].

Transistor scaling has been a fundamental approach toward increasing memory capacity, manufacturing low-power and high-performance devices [3] [4]. There have been numerous methods developed, involving improvements at the de-

This work is, supported in part by DST Nano Mission, Ministry of Electronics and Information Technology (MeitY), and Department of Electronics, through the Nano-electronics Network for Research and Applications (NNETRA) project of Govt. of India, and is funded by the German Bundesministerium für Wirtschaft (BMWI) and by the State of Saxony in the frame of the "Important Project of Common European Interest (IPCEI)".

Paritosh Meihar, Srinu Rowtu, Vivek Saraswat, Sandip Lashkare, Ajay Kumar Singh, and Udayan Ganguly are with the Department of Electrical Engineering, Indian Institute of Technology Bombay, Mumbai, 400076, India (e-mail: pari.1801@gmail.com, udayan@ee.iitb.ac.in).

Halid Mulaosmanovic, Stefan Dünkel, and Sven Beyer are with GlobalFoundries Fab1 LLC and Co. KG, 01109 Dresden, Germany.

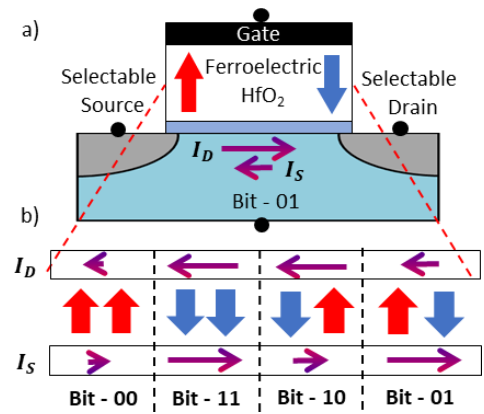

Fig. 1. **MirrorBit Concept:** a) FeFET based MirrorBit, and b) different polarization configurations and their corresponding drain ($I_D$) and source ($I_S$) currents.

vice level, to increase the memory capacity, such as dimensional scaling, 3D-stacking of flash transistors [5], developing new geometric designs like FinFET [6], and engineering material/gate-stack of emerging memories [7], [8]. In addition to physical alterations, there are programming methods which improve the storage on some specific memories, like Charge-trap flash [9] and FeRAM [10].

The ferroelectric memory, after the discovery of $HfO_2$-based ferroelectric devices in 2011 [11], has become a promising candidate and a competitor to the existing and other emerging memories [12]. The $HfO_2$-based ferroelectric memory offers fast switching speed, low operational voltage, high scalability, and CMOS compatibility [13]. We show MirrorBit operation in Ferroelectric-FET (FeFET). Unlike the localized charge trap phenomena in Charge-Trap Flash devices, FeFET works on the orientation of polarization of the ferroelectric layer. The conventional FeFET has two states or 1-bit of information, which correspond to saturated UP and DOWN polarizations [13]. A suitable biasing scheme creates a gradient in polarization, which gives rise to another bit of information (Fig. 1 (b)). The gradient in polarization affects the flow of the channel current in two opposite directions ($I_D$ and $I_S$), creating two threshold voltages ($V_T$) (Fig. 1 (c)). Hence, 4 states are created within the same devices effectively doubling the chip density.

TABLE I
TCAD MODEL PARAMETERS

| Parameter | Value | Unit |
|---|---|---|
| $P_r$ | 5.99 | $\mu C/cm^2$ |
| $P_s$ | 6 | $\mu C/cm^2$ |
| $E_c$ | 0.7 | $MV/cm$ |
| Interface traps concentration | $2.7 \times 10^{12}$ | $/cm^2$ |
| Gate length | 240 | nm |
| Gate width | 240 | nm |
| Channel doping | $3 \times 10^{16}$ | $/cm^3$ |
| $t_{HfO_2}$ | 10 | nm |
| $t_{SiO_2}$ | 0.5 | nm |

## II. METHODS

Fig. 2 (a) represents a device schematic of FeFET fabricated at GlobalFoundries' 28-nm bulk HKMG CMOS technology [12]. We have performed the electrical characterization, using Agilent B1530A 4-channel Waveform generator/Fast measurement unit (WGFMU), on FeFET devices of dimensions $L = 240$ nm and $W = 240$ nm. The pulse scheme involves a MirrorBit write and two read events. A TCAD model of the FeFET is developed and calibrated using experimental transfer curves to explain the origin and working of the MirrorBit. The TCAD model of the FeFET consists of a 10 nm HfO$_2$ layer along with an insulator layer (SiO$_2$) as gate oxide in the standard MOSFET model and ferroelectric properties are added to the HfO$_2$ layer, which makes the gate stack Metal-Ferroelectric-Insulator-Semiconductor (MFIS). The Preisach-based ferroelectric switching model is activated during the device simulation [14]. Preisach model alone is inadequate to capture the essence of the experimental transfer curves of the FeFET, as the fabricated FeFETs have several non-idealities. To fit the model appropriately, interface traps are also introduced. The model parameters are summarized in the table 1.

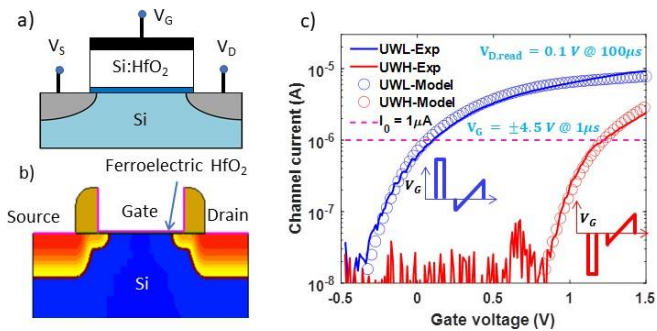

Fig. 2. **Experiment and Model calibration:** a) GF-FeFET Device schematic, b) 2D-model schematic of the FeFET in TCAD, and c) Experimental and simulated $I_D$ -$V_{GS}$. The model parameters are extracted by fitting the model with the experimental transfer curves. The Uniform-write low $V_T$ state (UWL) and Uniform-write high $V_T$ state (UWH) correspond to down and up polarization respectively. The inset shows the corresponding programming and read pulse schemes for UWL and UWH states.

## III. RESULTS AND DISCUSSION

### A. FeFET electrical characterization and TCAD model validation

The measured transfer characteristics of the FeFET (Fig. 2(c)) show two threshold voltages ($V_T$), which correspond to two conventional uniform polarization states. The uniform-write high $V_T$ (UWH) or Bit-00 state is written by applying $-4.5\,V$, $1\,\mu s$ voltage pulse at the gate and uniform-write low $V_T$ (UWL) or Bit-11 state by $4.5\,V$, $1\,\mu s$ voltage pulse.

Programming pulse width can be reduced up to few tens of nanoseconds as observed in [15]. Recently, it has been demonstrated that this kind of FeFET devices can even switch in the sub- nanosecond range, down to 300 ps [16], which however, necessitates a special probing equipment. For simplicity and due to our measurement set-up limitations, we adopt pulsewidths of $1\,\mu s$. We have chosen this time to ensure a full-scale saturated switching $V_T$.

To read the states, a ramp voltage, $V_G = -0.5\,V$ to $V_G = 1.5\,V$ with the read time of $100\,\mu s$, is applied at the gate, keeping drain voltage at $0.1\,V$ and source/substrate grounded. The $V_T$s of the device are extracted by constant current method at $I_0 = 1\mu A * W/L$.

To calibrate the TCAD model, a similar pulse scheme is used for both programming and reading the UW states. To write UWH state (starting with the unpolarized state), the voltage of the gate terminal is raised to $-4.5\,V$, which switches the polarization in the upward direction (Fig. 3 (a)). When the gate voltage is lowered back to $0\,V$, we observe that the conduction band (CB) energy is shifted up compared to the initial CB energy. This is a classic signature of non-volatile ferroelectric switching. Similarly, for UWL state, when the gate voltage is raised to $4.5\,V$ and reduced back to $0\,V$, we observe that the CB energy shifts downward (Fig. 3 (b)). Fig. 2 (c) shows an excellent match between the experimental and modelled transfer curves.

### B. MirrorBit write

The ferroelectric layer is divided into two parts (Fig. 3, device schematic), for ease in understanding and comparing the states. The programming of UWL and UWH states is explained in the previous section. Apart from the two conventional uniform-write states, additional states are programmed by generating a non-uniform electric field in the ferroelectric layer, which is achieved by applying a voltage at either source or drain terminal keeping other terminals grounded. If the voltage is applied at the drain, it is termed as the Drain Write (DW) state, and if the voltage is applied at the source, it is termed as the Source Write (SW) state.

In experiment, to program Source Write (SW) or Bit-01 state, the device is initialized with the UWL state. A voltage pulse $V_S = 3.6\,V$ (@ $400\,\mu s$) is applied at the source (keeping other terminals grounded) to obtain the SW state. The direction of the electric field near the source is from source to gate, which is opposite to the polarization, leading to a polarization switching near the source but not near the drain. Similarly, we write the DW state by initializing with UWL state followed by a $3.6\,V$ pulse at the drain.

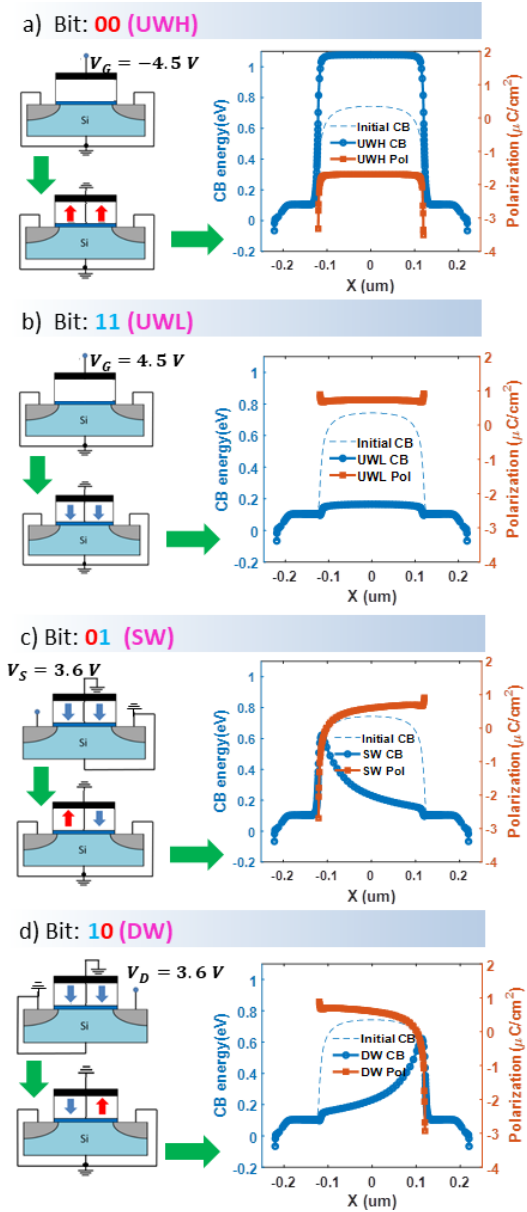

Fig. 3. **MirrorBit write:** a), b), c), and d) Band diagram and Polarization variation in the lateral direction in the channel and the ferroelectric HfO$_2$ layer respectively for UWH, UWL, SW and DW polarization configurations respectively. For uniform write case the energy band is uniformly shifted up or down, however, for DW/SW case the band is shifted up near terminal where the positive write voltage is applied, taking a triangular (Schottky barrier like) shape

In the TCAD, to understand the origin of MirrorBit SW and DW states, we apply the same pulse scheme as in the experiment. Starting with the initial UWL state (Fig. 3 (b)), SW (Bit-01) is obtained by applying $V_G = 0\ V$, $V_D = 0\ V$, $V_S = 3.6\ V$, and $V_{sub} = 0\ V$, producing a gradient in the electric field in the ferroelectric layer (Fig 3. (c)). This causes polarization to switch in a gradient fashion. After reducing the source voltage back to $0\ V$ we observe that the conduction band energy becomes triangular or Schottky-like barrier (Fig. 3 (c) blue curve). The peak of this barrier is towards the Source, where the write pulse was applied. Similarly, we obtain the DW (Bit-10) as well (Fig. 3 (d)).

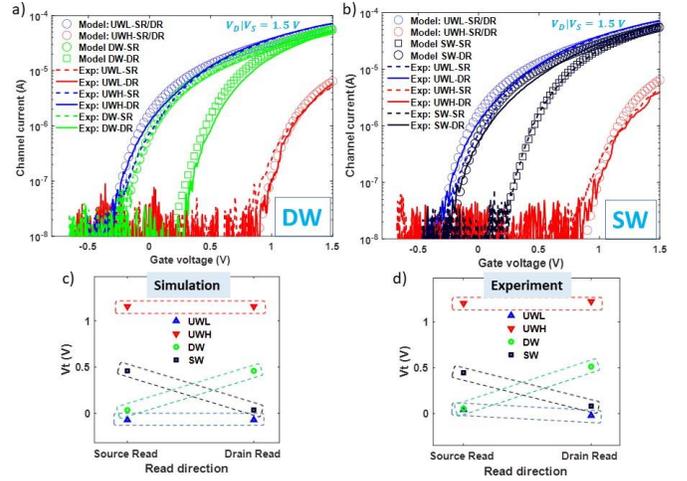

Fig. 4. **MirrorBit Memory:** a) and b) $I_D$-$V_G$ for UWL, UWH, DW, and SW cases showing that only one of the $V_T$ s shifts asymmetrically for DW/SW, but both $V_T$ s shift symmetrically for UW case, c) and d) corresponding $V_T$ distribution of all the states.

Although not explicitly considered in this work, with the lateral extension of the electric field and the formation of polarization gradient within the ferroelectric layer, a domain-to-domain interaction might be expected. This might be even more pronounced for different device geometries (like different channel lengths or widths) where percolation effects of specific domain alignments might play a role. These effects still need a deeper theoretical and experimental understanding.

### C. MirrorBit read

Reading the MirrorBit states involves measuring the channel currents in both the directions. When a read voltage is applied at the drain and $V_T$ is obtained through $I_D$-$V_G$, it is termed as Source Read (SR). Similary, it is called Drain Read (DR), if the read voltage is applied at the Source. In Fig. 4 (a) and (b), for UWH state, the $V_T$ s for both SR and DR are high and for UWL state, both $V_T$ s are low. This verifies the symmetric programming through uniform polarization switching. However, for SW (or DW) state, there is an asymmetric shift of $V_T$ s i.e. the SR (or DR) $V_T$ shifts to a higher value and the DR (or SR) $V_T$ remains nearer to the UWL $V_T$. The transfers curves are measured at a read voltage of $V_{read} = 1.5\ V$, and gate voltage is swept from $-0.5\ V$ to $1.5\ V$. Fig. 4 (c) and (d) summarize the SR/DR $V_T$s for UWH (Bit-00), UWL (Bit-11), SW (Bit-01), and DR (Bit-10) states. The experimental results have a close match with the simulated results.

During the calibration of the TCAD model, standard read voltage $V_D = 0.1\ V$ has been chosen. However, in experiment, it is observed that such low drain/source voltage is insufficient to produce any noticeable difference in the DR and SR currents for DW and SW states. Hence, a higher drain/source voltage of $1.5\ V$ is selected. This can also be understood from Fig. 5. When $V_G$ is kept fixed and source voltage (DR) is varied (Fig. 5 (a)), we observe an off-state current, similar to a Schottky barrier off-current. However, when the drain voltage (SR) is varied, we see a higher current which increases with read voltage. The reason for this asymmetric current lies in the shape of the conduction band energy, which resembles a Schottky-like barrier. The electron experiences two different barriers for SR and DR as shown in Fig. 5 (a) and (b). Hence, we observe a Schottky-like channel current (Fig. 5 (c)). Device-to-device variation of MirrorBit states is also measured for 6 devices (Fig. 6). We can observe that the states are clearly distinguishable.

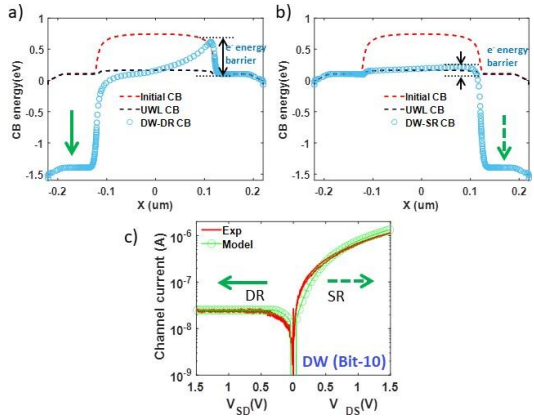

Fig. 5. **MirrorBit read:** a) $V_S$ and $V_D$ sweep for DW (Bit-10) showing the characteristics similar to a Schottky junction, b) and c) Band-diagram for DR and SR, showing the change in the peak heights, giving rise to different currents in two directions

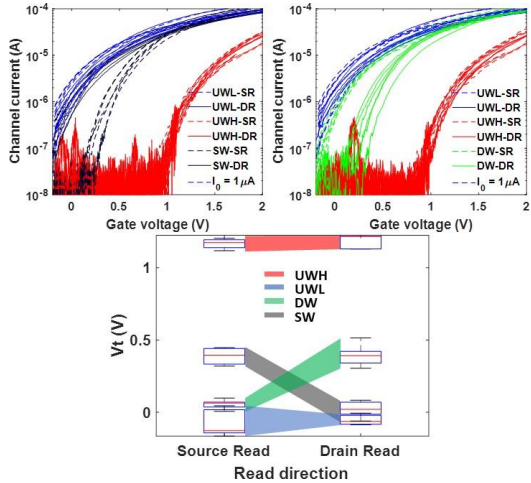

Fig. 6. **MirrorBit Device-to-device variation:** MirrorBit transfer curves and $V_T$ distribution measured for 6 devices.

### D. Retention and Endurance

We have measured the retention of DW and SW states at $25\ °C$ and $85\ °C$. After the DW state is programmed, SR and DR $V_T$s are measured over time. It can be observed, in Fig. 7, that all $V_T$s seem to converge caused by polarization reversal due to known effects such as depolarization field and charge trapping [15]. However, the memory window for the DW/SW is maintained till $10^5$ s. As expected, the convergence is faster at higher temperatures.

We have also perfomed the endurance test for all the states (Fig. 7). $V_T$s of all the states including the memory window are maintained for up to 1000 cycles. However, due to interface degradation mechanisms, the UWL $V_T$ starts shifting towards the UWH $V_T$ [12].

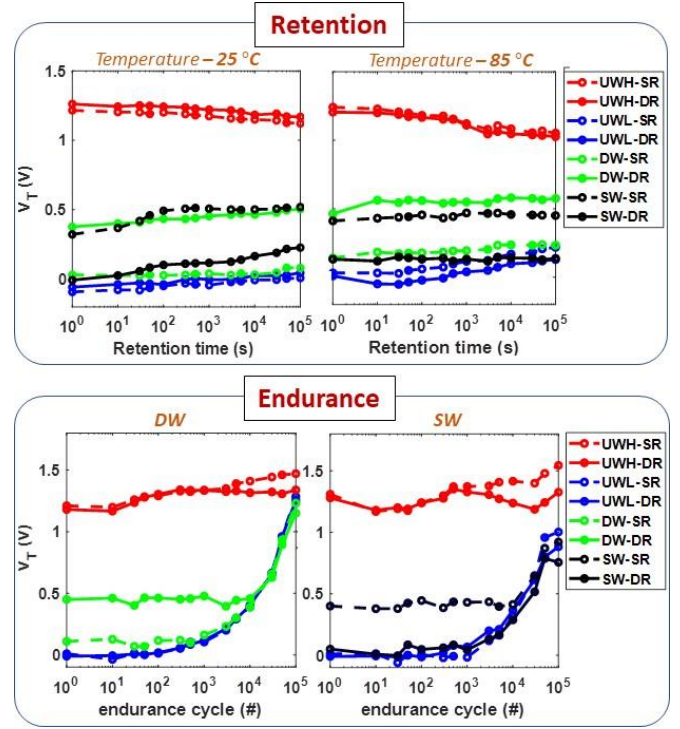

Fig. 7. **MirrorBit retention and endurance:** Retention of all the states is measured at $25\ °C$ and $85\ °C$. Endurance of DW and SW states including UWL and UWH shows good trend till 1000 cycles before degradation of UWL state starts dominating.

### E. MirrorBit array: Programming and sensing

Although, this work primarily focuses on the device demonstration of FeFET-based MirrorBit, we also propose, in brief, the array architecture with biasing and sensing schemes, which would require a separate study. Historically, the MirrorBit flash technology has preferred NOR array type for demonstrating successful functionality, scaling and high volume production [15]. Nevertheless, the FeFETs seem to show the best functionality in an AND-array type [16].

Fig. 8 (a) shows the AND-array architecture along with $V - V/2$ and $V - V/3$ programming schemes. The DW biasing scheme, for instance, shows appropriate applied voltages at word line (WL), source line (SL) and bit line (BL) terminals of the selected cell. However, there is a non-zero voltage drop in the half-selected and unselected cells. Although, the proposed programming schemes try to reduce the voltage levels which fall in the write disturb-free range in other cells, the accumulative [17] and minor loop switching [18] phenomena in ferroelectric devices may cause write-disturbs and/or accidental switching in other cells over repeated cycles [19]. According a recent study, however, it is shown that pulse-based biasing is disturb-free compared to continuous biasing [20]. Hence, due to debatable nature of this architecture and the pulse scheme, there is another architecture proposed with some tradeoffs.

To mitigate the write-disturbs, there is a select transistor connected to each terminal (Fig. 8 (b)). The select word line (SWL), select bit line (SBL), and select source line (SSL) control the access to WL, BL and SL, respectively. The SWL activates the corresponding row of the selected cell. Only for selected cell, the BL voltage is applied, keeping the gate and SL grounded. For other cells in this row, the BL and SL voltages are $0\ V$, causing no voltage drop in these devices. This architecture, surely, eliminates write-disturb issue, however, at the expense of increased area overhead at both cell and peripheral levels [21], [22].

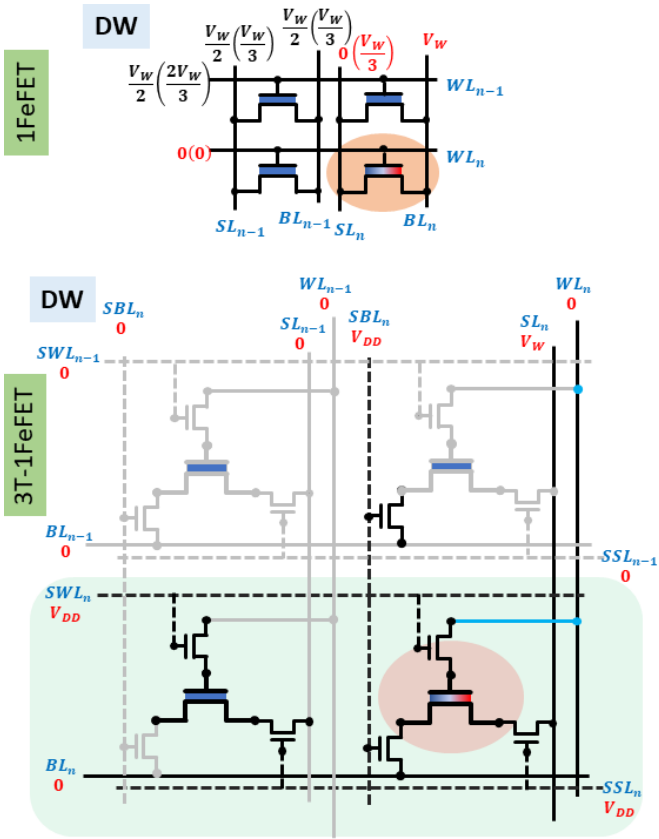

Fig. 8. **MirrorBit array:** 1FeFET array architecture has less area overhead but suffers from write disturbs for **V − V/2** and **V − V/3** programming schemes due to non-zero voltage drop in half-selected/unselected cells. 3T-1FeFET array architecture eliminates the write disturb by providing access transistor to each terminal at the cost of increased cell and peripheral area. The DW biasing scheme is shown as an example for both the architectures.

Conventionally, there are different sensing methods proposed for reading multibit cells [23], [24]. Sequential or serial sensing is the most appropriate read method for the MirrorBit memory, as two $V_T$s are required to be measured for determining a state. Fig. 9 represents the proposed serial sensing methodology. Considering 1FeFET architecture (for simplicity), the row decoder activates the corresponding row of the selected cell. The column decoder and sense configuration switches together select the appropriate lines depending on the type of read. For DR, for instance, the column decoder sets the SL to $V_{S,read}$ and connects BL to the sense amplifier (SA) through Sense Configuration Switches. Both current sensing and voltage sensing is possible in above architectures. The configuration will be swapped for SR. Due to distinguishable nature of $V_T$s, a single reference $V_T$ is sufficient for reading the states. If both sensed $V_T$s are above the reference $V_T$, then it's the UWH state. If DR $V_T$ and SR $V_T$ are higher and lower, respectively than reference $V_T$ then it's the DW state, and similarly other states can be sensed.

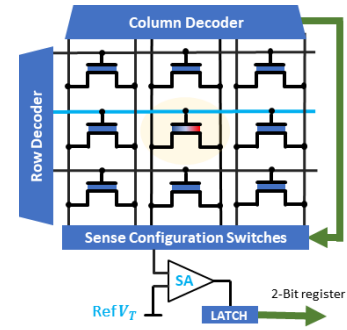

Fig. 9. **MirrorBit sensing block diagram:** Peripheral blocks are shown which implement serial sensing method on MirrorBit cell. The column decoder and sense configuration switches are configured simultaneously to select appropriate lines for SR and DR.

Hence, serial sensing gives area overhead advantage by using a single conventional SA at the expense of the increased sense time compared to parallel sensing which is faster but requires 3 SAs, thereby increasing the area overhead of the peripherals [23]. One latch will suffice due to sequential nature of the read, and the sensed state is transferred to 2-Bit shift register. Therefore, in terms of SA's design and read speed trade-off, there are no additional requirements and considerations for ferroelectric MirrorBit compared to the already existing multibit sensing techniques.

## IV. CONCLUSIONS

We showed the operation and working of a FeFET-based MirrorBit, experimentally and through simulation. The device-to-device variation shows a tight and distinguishable distribution of $V_T$s. The simulation reveals a Schottky-like triangular barrier formation, which explained the asymmetry in the channel current in two directions. The asymmetric states DW and SW are retaining their states for $> 10^4$ s. Lastly, the array implementation along with memory biasing scheme is also proposed. Hence, without any process alteration, purely through biasing, the density of bits has been doubled. This opens up avenues for using fundamental device characteristics beyond its defined functionality.